\documentstyle[amssymb,preprint,aps]{revtex}

\begin{document}
\draft
\author{M. Jaime\thanks{%
Present address: Los Alamos National Laboratory, MSK764, Los Alamos, NM 87545%
}, P. Lin, M. B. Salamon and P.\ D. Han}
\address{Department of Physics and Materials Research Laboratory\\
University of Illinois at Urbana-Champaign\\
1110 W. Green Street, Urbana, IL 61801}
\title{Low-temperature electrical transport and double exchange in La$_{0.67}$%
(Pb,Ca)$_{0.33}$MnO$_3$.}
\date{
\today%
}
\maketitle

\begin{abstract}
The resistivity in the ferromagnetic state of flux-grown La$_{2/3}$(Pb,Ca)$%
_{1/3}$MnO$_3$ single crystals, measured in magnetic fields up to 7 T,
reveals a strong quadratic temperature dependence at and above 50 K. At
lower temperatures, this contribution drops precipitously leaving the
resistivity essentially temperature independent below 20 K. The Seebeck
coefficient also reflects a change of behavior at the same temperature. We
attribute this behavior to a cut-off of single magnon scattering processes
at long wavelengths due to the polarized bands of a double-exchange
ferromagnet.
\end{abstract}


The substitution of divalent A-atoms in such manganites as La$_{1-x\text{ }}$%
A$_x$MnO$_3$ has long been known to induce ferromagnetism and anomalous
transport properties, \cite{early} yet an understanding of the underlying
mechanisms remains incomplete. The double exchange mechanism, first proposed
by Zener\cite{zener} and later developed by Anderson and Hasegawa \cite{AH},
by Kubo and Ohata (KO) \cite{KO} and by Furukawa \cite{furu}, is generally
agreed to provide a description of the ferromagnetic ground state. In that
model, strong Hund's Rule coupling enhances the hopping of $e_{g\text{ }}$%
electrons between neighboring Mn$^{3+}$ and Mn$^{4+}$ ions by a factor $\cos
(\theta _{ij}/2)$, where $\theta _{ij}$ is the angle between the spin of
their respective $t_g$ cores, thereby producing a ferromagnetic interaction.
The onset of ferromagnetic order either shrinks the minority-spin bandwidth
to zero as the majority bandwidth grows (KO picture) or shifts spectral
weight from minority- to majority-band without a change in bandwidth or
position (Furukawa picture). Both are distinct from the usual intinerant
ferromagnetic scenario in which there is simply a rigid energy displacement
of majority and minority bands. It has been argued \cite{millis} that
electron-phonon coupling is important, but we assume that those effects
dominate near and above the ferromagnetic transition which is not our focus
here.

In the KO picture, the minority band narrows at absolute zero to a resonance
at the band center, resulting in an energy gap $\mu $ between the majority
Fermi level and the lowest accessible minority-spin state. Because single
spin-wave scattering requires a spin flip, KO argue that the characteristic $%
T^2$ temperature dependence of the resistivity of metallic ferromagnets will
be suppressed by a factor $\exp (-\mu /k_BT)$. Extending the standard
perturbation calculation of Mannari \cite{mannari} to consider two-magnon
processes, they predict a leading $T^{9/2}$ temperature dependence in the
resistivity. \ However, a $T^2$ dependence of the resistivity of manganites
is typically observed and has been attributed to electron-electron
scattering. \cite{urushi,snyder} In this paper we present new resistivity
data on single crystals, and demonstrate that the quadratic temperature
dependence is strongly suppressed as the temperature is reduced, becoming
undetectable below 10 K. We argue that this contribution is due to
single-magnon processes and that our results support the gradual development
of a half-metallic state. However, the disappearance of minority spin states
at the Fermi energy appears to be less abrupt than in the KO model, favoring
instead Furukawa's, in which the minority-spin states reappear at the Fermi
surface as soon as the magnetization decreases from its fully ordered value.
Band structure calculations \cite{pickett} also indicate that minority spin
states persist at $E_F,$ even at T = 0 K. To explore the magnon cut-off, we
have extended Mannari's model to a situation in which a minimum magnon
energy is required to induce spin-flip transitions. This is essentially the
argument leading to the suppression factor predicted by Kubo and Ohata.

High quality single crystals of nominal composition La$_{0.66}$(Pb$_{0.67}$Ca%
$_{0.33}$)$_{0.34}$MnO$_3,$ determined by inductively coupled plasma
spectroscopy on samples from the same batch, were grown from a 50/50 PbF$_2$%
/PbO flux.\cite{han} Samples grown without Ca in the flux exhibited low
Curie temperatures, indicating poor sample quality and/or deviations from
optimal doping. Specimens were cut along crystalline axes from larger
pre-oriented crystals. X-ray diffractometry shows a single
pseudo-orthorhombic structure with lattice parameters $a=5.472(4)\;\AA
,\;b=5.526(6)\;\AA ,\;$and $c=7.794(8)\;\AA .$ Gold pads were evaporated
onto both oriented and unoriented crystals using both standard four-terminal
and Montgomery eight-corner contact arrangements. Electrical contact was
made with Au leads attached with silver paint. After a 2-hour anneal at 300 $%
^{\circ }C,$ contact resistances $<1\;\Omega $ were achieved. The Seebeck
coefficient was measured using a variation of the two-heater method
described by Resel, et al. \cite{resel}; Chromel and Constantan wires,
calibrated against a YBa$_2$Cu$_3$O$_7$ superconducting sample, permitted
this method to be used up to 80 K. The resistivity was measured from 2 K to
400 K using either a standard four-terminal dc method, or a dc resistance
bridge in a Quantum Design PPMS system. Magnetization data on the same
crystals were obtained using a Quantum Design MPMS magnetometer.

Figure 1 shows the resistivity of sample sc3, a single crystal of dimensions
1.04$\times $1.24$\times $0.3 mm$^3$ with $T_C=300$ K, vs the square of the
temperature in fields up to 70 kOe. The data show a dominant $T^2$
temperature dependence with evidence of a small $T^5$ contribution (10 $\mu
\Omega $ cm at 100 K). A calculation of the $T^{9/2}$ contribution predicted
by KO for two-magnon processes predicts only 0.5 $\mu \Omega $ cm at 100 K
with appropriate parameters. It is likely, then, that this is the usual $T^5$
contribution from electron-phonon processes. The inset to Fig. 2 shows that
the data do not follow a $T^2$ dependence to the lowest temperatures.
Rather, they deviate gradually from the curve $\rho _0+\alpha (H)T^2,$ fit
over the range $60\leq T\leq 160$ K, saturating at an experimental residual
resistivity $\rho _0^{exp}=91.4$ $\mu \Omega $\ cm, comparable to values
observed by Urushiba et al. \cite{urushi}, but $\sim $7\% larger than $\rho
_0$ (see Table I). This conclusion is not changed by including the $T^5$
contribution. Fits to data taken in various fields show that $\alpha _H$
decreases with increasing field and is the source of the small negative
magnetoresistance at low temperatures. To quantify the disappearance of the $%
T^2$ contribution, we numerically differentiate the data, plotting $\alpha
(H)^{-1}d\rho /d(T^2)$ in Fig. 2. We have not subtracted the $T^5$
contribution which gives a slight upward curvature to the data at higher
temperatures.

Previous investigators have attributed the $T^2$ term in the resistivity to
electron-electron scattering, as expected for conventional Fermi liquids.
The relaxation rate, however, is then of order $(k_BT)^2/\hbar E_F.$ \cite
{ashcroft} With an electron density of $5.7\times 10^{27}$ m$^{-3}$ (1/3
doping) and an effective mass ratio $m^{*}/m=2.5$ \cite{mstar} we find $%
E_F\simeq 0.5$ eV and the $e-e$ relaxation rate of order $2\times 10^{11}$ s$%
^{-1}$ at 100 K. The observed $T^2$ contribution at that temperature (cf.
Table I) is 100 $\mu \Omega $ cm which with the same parameters corresponds
to a relaxation rate of $6\times 10^{13}$ s$^{-1},$ more than two orders of
magnitude larger. Rather than vanishing, what is more, $e-e$ scattering
should become more apparent as the temperature is reduced. We conclude that $%
e-e$ scattering is an unlikely explanation for the observed quadratic
dependence on temperature.

To test whether single-magnon scattering provides a viable mechanism, we
have extended the usual calculation of the electron-magnon resistivity \cite
{mannari} to allow the minority-spin sub-band to be shifted upward in energy
such that its Fermi momentum differs by an amount $q_{min}$ from that of the
majority sub-band. This should be a reasonable approximation in the
intermediate temperature regime in which both minority and majority bands
have substantial densities of states at $E_F.$ The one-magnon contribution
can then be written as $\rho _\epsilon (T)=\alpha _\epsilon T^2,$ where 
\begin{equation}
\alpha _\epsilon =\frac{9\pi ^3N^2J^2\hslash ^5}{8e^2E_F^4k_F}\left( \frac{%
k_B}{m^{*}D}\right) ^2I(\epsilon ).  \label{eq1}
\end{equation}
In this equation, $NJ$ is the electron-magnon coupling energy which is large
and equal to $\mu =W-E_F$ in the double-exchange Hamiltonian of KO; $2W$ is
the bandwidth. The magnon energy is given by $Dq^2,$ and we have defined 
\begin{equation}
I(\epsilon )=\int_\epsilon ^\infty \frac{x^2}{\sinh ^2x}dx.  \label{eq2}
\end{equation}
The lower limit is $\epsilon =Dq_{min}^2/2k_BT,$ where $Dq_{min}^2$ is the
minimum magnon energy that connects up- and down-spin bands; this result
reproduces Mannari's calculation in the limit $\epsilon \rightarrow 0,$ and
KO's exponential cut-off for large $\epsilon .$ Within the spin-wave
approximation, the low temperature magnetization is given by $%
M(T)=M(0)-BT^{3/2}-...,$ where $B=0.0587g\mu _B(k_B/D)^{3/2}.$ The stiffness
constant $D$ has been determined by neutron scattering \cite{lynn,baca} and
muon spin resonance \cite{heff} to be $D\approx 135-170$\ meV \AA $^2.$ The
inset to Fig. 1 shows the magnetization for this sample, from which we
extract $B(10\;$kOe$)$ (Table I) and the value $D=165$ meV \AA $^2,$ in good
agreement with other results. The effect of an applied field is to open a
gap $\Delta =g\mu _BH$ in the magnon spectrum, resulting in a
field-dependent coefficient $B(H)\approx B(0)(1-1.36\sqrt{\Delta /k_BT}+...).
$ The same gap enters the hyperbolic function in Eq.(2), with the result
that $\delta \alpha \equiv \alpha _0(0)-\alpha _0(H)\varpropto \Delta /k_BT.$
Using our high-temperature ($\epsilon \approx 0)$ fits, we plot $B(H)$ vs $%
\sqrt{\delta \alpha }$ in Fig. 3; that these are proportional supports our
assertion that the $T^2$\ contribution is due to single-magnon scattering.\
The zero-field intercept corresponds to $D$ = 140 meV \AA $^2,$\ reasonably
close to the value reported by Perring, et al. for Pb-doped samples. \cite
{perring}

The Seebeck coefficient $S(T)$ provides additional information on the nature
of transport at low temperatures. Fig. 4 shows $S(T)$, measured on the same
sample at $H=0$ and $80$ kOe. At the lowest temperatures, $S(T)$ is
positive, linear in temperature, and extrapolates to zero as $T\rightarrow 0$%
. The field dependence is small and negative. The large slope suggests, from
the Mott formula, that the resistivity is a strong function of energy at $%
E_{F\text{.}}$ There is a sharp deviation from linear behavior in the
temperature range in which the $T^2$ -dependence of the resistivity becomes
dominant and the field dependence changes sign and becomes larger. This
presumably arises from the onset of magnon scattering which, being a spin
flip process, must involve the minority spin band, and which therefore has a
different dependence on energy near $E_F.$ We note that the peak in the low
temperature thermopower that is regularly seen in thin-film samples \cite
{tep} is absent here, and is therefore not intrinsic to these materials.

At high temperatures, the lower limit of the integral in Eq. (\ref{eq1}) can
be set equal to zero, leaving only the coupling energy $NJ=W-E_F$ as a
parameter. Equating the calculated value to the experimental $\alpha $ from
Table I fixes the coupling to be $W-E_F$ $\approx 1.0$ eV or $W\simeq 1.5$
eV, in good agreement with a virtual crystal estimate of the band width. 
\cite{pickett2} In Fig. 2 we have plotted $\alpha _0^{-1}d(\alpha _\epsilon
T^2)/d(T^2)$ assuming $D(0)q_{min}^2=4$ meV and including the temperature
dependence observed experimentally, $D(T)/D(0)=(1-T/T_C)^{0.38}$ \cite{baca}
which is important only at higher temperatures. While the curve follows the
data qualitatively, it is clear that the minimum magnon energy is
substantially larger than 4 meV at low temperatures, and decreases rapidly
with increasing temperature. We note that the leading correction to a $T^2$
dependence from Eq.(\ref{eq1}) is negative and linear in $T$, as observed
elsewhere. \cite{lofland} 

Clearly, our extension of the magnon resistivity calculation to spin-split
parabolic bands greatly oversimplifies the changes in the minority-spin band
that accompany magnetic ordering. As a consequence, our calculation cannot
be expected to represent accurately the cut-off of magnon scattering due to
loss of minority-spin phase space. Nonetheless, the rapid suppression of the 
$T^2$ contribution to the resistivity and the agreement between its
magnitude at higher temperatures with parameters expected for the manganites
confirm the basic picture. In the intermediate temperature regime, $0.2\leq
T/T_C\leq 0.5$ here, the manganites appear to be normal metallic
ferromagnets with the resistivity dominated by spin-wave scattering. At
lower temperatures, the increasingly half-metallic character of the material
is manifested by a temperature dependent cut-off of the spin-wave scattering
process, leaving in its wake only residual resistivity from the intrinsic
doped-in disorder and indistinguishable phonon and two-magnon contributions.
As these heavily doped materials have significant disorder and large
residual resistivities we recall that strongly disordered materials also
exhibit $T^2$ regimes below half the Debye temperature. \cite{nagel} This
result, however, an extension of the Ziman theory of liquid metals, sets an
upper limit of $\alpha T^2/\rho _0\simeq 0.03$ before the resistivity
changes to a linear temperature dependence; our ratio is unity at 100 K with
no evidence for a linear regime. We conclude that the quadratic temperature
dependence is not due to phonon scattering in a strongly disordered material.

In summary, our low temperature resistivity data on well-characterized
single crystal samples exhibit a dominant $T^2$ temperature dependence above
50 K that vanishes abruptly at lower temperatures and is consistent in
magnitude and magnetic field dependence with one-magnon scattering
processes. Our data cannot be explained by electron-electron scattering as
proposed in previous reports.\cite{tep,urushi,snyder} While oversimplified,
our extension of the standard calculation of one-magnon resistivity to
account for spin-split bands gives a qualitative account of the
half-metallic suppression of spin-wave scattering at low temperatures. The
onset of significant magnon-mediated scattering near 30 K is also reflected
in a change in both the temperature and field dependence of the thermopower.
While two-magnon processes may contribute, these are expected to be very
small, and cannot be separated from ordinary phonon scattering nor from the
rapid increase in resistance that occurs near the Curie temperature. Our
data demonstrate that the half-metallic ground state of the doped manganites
quickly evolves with increasing temperature into a more conventional,
metallic ferromagnet.

We have benefited from technical assistance from Dr. S-H. Chun and from
discussions with Dr. H. R\"{o}der. This work was supported by the Department
of Energy, Office of Basic Energy Sciences through Grant No.
DEFG02-91ER45439 at the University of Illinois and by National Science
Foundation Grant No. DMR-9120000 through the Science and Technology Center
for Superconductivity.

\begin{table}[tbp] \centering%
\begin{tabular}{|l|c|c|c|c|}
\hline\hline
La$_{2/3}$A$_{1/3}$MnO$_3$ & $\rho _0$ & $\alpha (H)$ & $B(H)$ & $%
l_{mag}^{77K}$ \\ 
& $\mu \Omega $ cm & $n\Omega $ cm K$^{-2}$ & Gauss K$^{-3/2}$ & $\AA $ \\ 
\hline\hline
{\normalsize A = Ca, Pb } &  &  &  &  \\ 
sample La5 &  &  &  &  \\ 
H = 0 & $133.3\underline{3}$ & $12.7\underline{9}$ & - & $53$ \\ 
10 kG & - & - & $0.01\underline{4}$ & - \\ \hline\hline
{\normalsize A = Ca, Pb } &  &  &  &  \\ 
sample SC3 &  &  &  &  \\ 
H = 0 & $85.\underline{0}$ & $9.9\underline{0}$ & - & $76$ \\ 
10 kG & $86.\underline{8}$ & $9.7\underline{3}$ & $0.013\underline{2}$ & $76$
\\ 
30 kG & $86.\underline{8}$ & $9.6\underline{8}$ & $0.012\underline{5}$ & $77$
\\ 
50 kG & $85.\underline{1}$ & $9.5\underline{7}$ & $0.011\underline{3}$ & $79$
\\ 
70 kG & $86.\underline{2}$ & $9.4\underline{6}$ & $0.010\underline{9}$ & $78$
\\ \hline\hline
\end{tabular}
\caption{Fitting parameters for resistivity and magnetization as described in the text. 
The mean free path from electron magnon scattering at 77 K has been calculated.}%
\end{table}%

\begin{figure}
\caption{The resistivity vs $T^2$ for magnetic fields up to 70 kG in sample sc3.  Inset:  
The difference between the magnetization and its zero-temperature extrapolation vs
 temperature on a log-log plot, showing $T^{3/2}$ and $T^{5/2}$ contributions}
\label{fig1 }
\end{figure}%

\begin{figure}
\caption{The numerical derivative of the data, 
$\alpha(H)^{-1}d\rho/d(T^2)$, for zero field ($\blacksquare $) and 70 kG ($\square$).  
The dashed curve is the same quantity calculated for the magnon model with
 $Dq_{min}^2 /k_B =44$  K. Inset: raw data  minus
calculated residual resistivity ($\blacksquare $) and least-squares curve (dotted line) vs 
temperature. Note the violation of Matthiessen's rule at low temperatures. }
\label{fig.2 }
\end{figure}%

\begin{figure}
\caption{The magnetization coefficient $B$ vs the resistivity coefficient 
$[\alpha (H=0) - \alpha (H)]^{1/2}$ for four different fields.  Proportionality is 
expected if the $T^2$  temperature dependence is due to one-magnon processes. 
The line is a linear fit to the data.}
\label{ fig.3}
\end{figure}%

\begin{figure}
\caption{The Seebeck coefficient S vs temperature.  It is positive and metal-like at low
temperature, has an anomalous kink near 30 K, and develops a positive field dependence
above 40 K.  The dashed line is a linear fit in the low temperature regime. }
\label{fig.4 }
\end{figure}%


\begin{references}
\bibitem{early}  G.H. Jonker and J.H. Van Santen, Physica {\bf 16}, 337
(1950); J. Volger, Physica XX, 49 (1954).

\bibitem{zener}  C. Zener, Phys. Rev., {\bf 82, }403 (1951).

\bibitem{AH}  P.W. Anderson and H. Hasegawa, Phys. Rev. {\bf 100}, 675
(1955).

\bibitem{KO}  K. Kubo and N. Ohata, J. Phys. Soc. Jpn. {\bf 33}, 21, (1972).

\bibitem{furu}  N. Furukawa, J. Phys. Soc. Jpn. {\bf 64}, 3164 (1995);{\it \
ibid}. 2734, (1995); {\it ibid.} {\bf 63}, 3214 (1994).

\bibitem{millis}  A.J. Millis, P.B. Littlewood, and B. Shraiman, Phys. Rev.
Lett. {\bf 74}, 5144 (1995).

\bibitem{mannari}  I. Mannari, Prog. Theor. Phys. Jpn. {\bf 22}, 335 (1959).

\bibitem{urushi}  A. Urushibara et al., Phys. Rev. B {\bf 51}, 14103 (1995).
Also S.E. Lofland et al., Phys. Rev. B {\bf 52}, 15058 (1995).

\bibitem{snyder}  P. Schiffer et al., Phys. Rev. Lett. {\bf 75}, 3336
(1995). Also G.J. Snyder et al., Phys. Rev. B {\bf 53}, 14434 (1996).

\bibitem{pickett}  W. E. Pickett and D. J. Singh, Phys. Rev. B {\bf 53},
1146 (1996); D. J. Singh and W. E. Pickett (to be published).

\bibitem{han}  P.D. Han and D.A. Payne, J. Mater. Res. (submitted).

\bibitem{resel}  R. Resel et al. Rev. Sci. Inst. {\bf 67}, 1970\ (1996).

\bibitem{ashcroft}  N. Ashcroft and D. Mermin, {\sl Solid State Physics}
(Holt, Rinehart, and Winston, New York, 1976) p. 348

\bibitem{mstar}  J.J. Hamilton et al. Phys. Rev. B {\bf 54}, 14926 (1996).

\bibitem{lynn}  J.W. Lynn et al., Phys. Rev. Lett.{\bf 76}, 4046 (1996).

\bibitem{baca}  J. Fernandez-Baca et al., (preprint, 1997).

\bibitem{heff}  R.H. Heffner et al., Phys. Rev. Lett., {\bf 77}, 1869 (1996).

\bibitem{perring}  T.G. Perring et al., Phys. Rev. Lett. {\bf 77}, 711
(1996).

\bibitem{tep}  M. Jaime et al., Appl. Phys. Lett. {\bf 68}, 1576 (1996); M.
Jaime et al., Phys. Rev. B {\bf 54,} 11914 (1996); M.F. Hundley and J.J.
Neumeier, Phys. Rev. B {\bf 55}, 11511 (1997).

\bibitem{pickett2}  W. E. Pickett and D. J. Singh, Phys. Rev. B {\bf 55},
R8642 (1997).

\bibitem{lofland}  S.E. Lofland, et al. Phys. Rev. B {\bf 52}, 15 058 (1995).

\bibitem{nagel}  S. R. Nagel, Phys. Rev. B {\bf 16}, 1694 (1977).
\end{references}
\end{document}